\documentclass[aps,pre,twocolumn]{revtex4-2}
\usepackage{amsmath}
\usepackage{amsfonts}
\usepackage{graphicx}
\usepackage[font=small,skip=0pt]{caption}
\usepackage{commath}
\usepackage{caption}
\usepackage{tabularx,ragged2e,booktabs,caption}
\usepackage{xcolor}
\usepackage{graphicx}
\usepackage{har2nat}

\newcolumntype{C}[1]{>{\Centering}m{#1}}

\makeatletter
\newcommand{\vast}{\bBigg@{4}}
\newcommand{\Vast}{\bBigg@{5}}
\makeatother

\begin{document}

\begin{abstract}
In order to understand controlling a complex system, an estimation of the required effort needed to achieve control is vital. Previous works have addressed this issue by studying the scaling laws of energy cost in a general way with continuous-time linear dynamics. However, continuous-time linear dynamics is unable to capture conformity behavior, which is common in many complex social systems. Therefore, to understand controlling social systems with conformity, discrete-time modelling is used and the energy cost scaling laws are derived. The results are validated numerically with model and real networks. In addition, the energy costs needed for controlling systems with and without conformity are compared, and it was found that controlling networked systems with conformity features always requires less control energy. Finally, it is shown through simulations that heterogeneous scale-free networks are less controllable, requiring a higher number of minimum drivers. Since the conformity-based model relates to various complex systems, such as flocking, or evolutionary games, the results of this paper represent a step forward towards developing realistic control of complex social systems. 
\end{abstract}
\title{Energy cost study for controlling complex social networks with conformity behavior}
\author{Hong \surname{Chen}}
\author{Ee Hou \surname{Yong} }
\email{Corresponding author: eehou@ntu.edu.sg}
\affiliation{Division of Physics and Applied Physics, School of Physical and Mathematical Sciences, Nanyang Technological University, Singapore, 637371, Singapore. }
\maketitle

\section{Introduction}
The controllability of complex networks have been studied extensively in recent years \cite{liu2011controllability,liu2016control,wang2012optimizing,cowan2012nodal,sun2013controllability,gao2014target,
iudice2015structural,whalen2015observability,posfai2016controllability,liu2017controllability,klickstein2017locally,li2018enabling,
zhao2019controllability,jiang2019irrelevance}. Owing to the ubiquity of networked systems in social \cite{albert2002statistical}, biological \cite{rajapakse2011dynamics,yan2017network,kenett2018computational}, ecological \cite{zhang2020co}, technological \cite{ruths2014control}, or financial \cite{delpini2013evolution} systems, understanding how to control them is important. The nodes of a network represent individual members, for example, species in an ecological network \cite{albert2002statistical} or individuals in an opinion network \cite{castellano2009statistical}, and links between nodes represent coupled interactions, for example, when a particular specie preys on the other, or when the opinion of an individual influences the opinion of others. Based on their complex networked interactions, complex dynamical systems could be modelled with state equations, where the states represent, for example, population level or support/opposition of a particular idea. Left to their own devices, these complex dynamical systems would evolve over time in a certain way. However, by introducing appropriate external interventions, the node states evolution could be altered (controlled) and made to behave in some other prescribed ways \cite{liu2011controllability}. Therefore, by finding out which nodes in the network need to receive the appropriate external control signals (nodes which are injected with control signals are called driver nodes)  \cite{liu2011controllability,yuan2013exact}, a complex system could be controlled and have its node states steered towards the desired final state vector.

Within the literature, particular attention has been paid towards the energy cost needed for control. The energy cost relates to the amount of effort that the drivers have to consume to steer the state vector, and is affected by the inverse of the Gramian \cite{yan2012controlling}, which is dependent on the network structure, driver nodes, and control time $T_f$. Numerically, the energy cost has been studied in the context of optimization \cite{li2015minimum,lindmark2018minimum,ding2019key,chen2020optimizing} and computation \cite{chen2016energy,wang2017physical}. Analytically, scaling behaviors have been derived in terms of number of drivers \cite{yan2015spectrum}, target nodes cardinality \cite{klickstein2017energy}, and $T_f$ \cite{yan2012controlling,duan2019energy,duan2019target}. 

While the aforementioned works have been adequate in their treatment of network control, the results presented have so far generalised the network dynamics to be continuous-time and linear, without taking into account conformity. Given that networked controllability is restricted to the study of networks with linear dynamics owing to the difficulty in controlling nonlinear dynamical systems \cite{liu2016control}, continuous-time linear dynamics modelling is unable to capture conformity, which are often nonlinear \cite{madeo2020self}, the same way that discrete-time linear dynamics modelling can \cite{wang2015controlling}. As shown in Ref.\ \cite{wang2015controlling}, when the dynamics of the networked system has conformity behavior, where each member adapts and mimics its nearest neighbors' node states, the results obtained can differ when comparing network dynamics with and without conformity. For example, the eigenvalues of unweighted chain, ring, and complete graphs differ when comparing the systems with and without conformity. Besides, the conclusions drawn from structural controllability using a generalised network structure were inconsistent when taking into account self-dynamics \cite{cowan2012nodal}. Further, conformity dynamics models a significant class of social systems in a more realistic way, which a generalised linear dynamics may fail to capture. For example, evolutionary games \cite{wu2014role}, learning behaviors \cite{van2013potent, whiten2005conformity} and collective movements \cite{vicsek2012collective,vicsek1995novel,nagy2010hierarchical,ward2008quorum,buhl2006disorder}
 of animals are all described by models with conformity feature. Given the sensitivity of model specificities and the importance of control energy, it would be worthwhile to revisit previous calculations, keeping in mind conformity dynamics.

In this paper, the energy cost needed to control a complex network with conformity behavior is studied. The idea behind this work is to check what are the effects of the mechanism of conformity on the energy cost. To do so, the energy cost is studied analytically through a series of scaling laws. In general, the energy cost is affected by several different parameters such as initial state vector, final state vector, final control time $T_f$, number of drivers, and even network eigenvalues. Therefore, a systematic study of the analytical scaling laws of the control energy is advantageous, as compared to numerical ones, because it characterizes exactly the boundaries of the energy cost based on these different parameters and allows the energy cost to be estimated. Based on the research results, it was found that for controlling networks with discrete-time linear dynamics and conformity behavior, the lower and upper bound energy cost scales as a function of $T_f$ similarly to those of continuous-time linear dynamics without conformity behavior. While their scaling behaviors are similar, their energy costs are not. When comparing the energy cost needed to control systems with and without conformity, it was found that the energy cost of those with conformity features are always lower, suggesting that the mechanism of conformity is beneficial to the control of networked systems. Finally, the controllability of scale-free (SF) networks with degree-distribution $P(k) \sim k^{-\gamma}$ was also studied, and it was found that with or without conformity, the scaling exponent $\gamma$ plays an important role in controllability, where SF networks with lower scaling exponent $\gamma$ are less controllable, requiring a higher number of minimum driver nodes.  

\section{Problem formulation} \label{Problem formulation}
In networked systems with conformity behavior, each node $i$ mimics the strategies (or opinions) of its nearest neighbors $j$ by adapting its node state in the next round as the average of its $n_i$ nearest neighbors' node states in the current round \cite{wang2015controlling}: 
\begin{equation} \label{conformity behavior dynamics}
x_i(\tau +1)=\frac{1}{s_i}\sum_{j=1}^{n_i} \hat{a}_{ij} x_j (\tau),
\end{equation} 
where $x_i(\tau)$ is the strategy of node $i$ in discrete time $\tau$, $s_i=\sum\limits_{j=1}^N \hat{\bf A}_{ij}$ is the weighted strength of node $i$, with $\hat{\bf A}=\{\hat{a}_{ij}\}$ being the $N \times N$ network matrix, where $\hat{a}_{ij}$ is the directed weighted connection pointing from node $j$ to node $i$, and zero otherwise. Note that $x_i$ is a continuous variable that describes, for example, the degree of opinion in an opinion network \cite{acemoglu2010spread,liu2016control}, where a high $x_i$ value indicate support for opinion $A$, and a low $x_i$ value indicate support for opinion $B$. In evolutionary games, $x_i\in [0,1]$ denotes the probability of node $i$ to select a particular strategy. For example, $x_i=1$ indicates that node $i$ always chooses to cooperate, while with $x_i=0$, node $i$ always defects \cite{wang2015controlling}. Eqn.\ (\ref{conformity behavior dynamics}) belongs to a class of conformity-based memory-one strategy games \cite{perc2013evolutionary,perc2010coevolutionary}, where the strategy played by an agent is influenced only by the strategies of their immediate neighbors in the round prior. For example, a fair player has a weighted probability to cooperate with other players based their immediate neighbors' strategies played in the previous round \cite{hilbe2014cooperation}. 

Introducing input control signal terms to (\ref{conformity behavior dynamics}), and writing the state equation in vector notation, the complex system has the following discrete linear time-invariant (LTI) dynamics \cite{wang2015controlling}:
\begin{eqnarray} \nonumber \label{conformity behavior vectorised}
{\bf x}(\tau +1) =& {\bf S}^{-1}\hat{\bf A}{\bf x}(\tau) + {\bf B}{\bf u}(\tau)\\*
=&{\bf A} {\bf x}(\tau) + {\bf B}{\bf u}(\tau), 
\end{eqnarray}\\
where ${\bf x}(\tau)=[x_1(\tau), x_2(\tau), ..., x_N(\tau)]^T$ is the state vector which captures the node states, ${\bf u}(\tau)=[u_1(\tau),u_2(\tau), ..., u_M(\tau) ]^T$ is the input control signals which attach to the complex network to alter the node states, ${\bf B}=\{b_{ij}\}$ is the $N \times M$ input matrix which tracks where the $M$ number of control signals are placed, where $b_{ij}=1$ if control signal $j$ attaches to node $i$ (nodes which receive control signals are called driver nodes), and $b_{ij}=0$ otherwise, ${\bf S}^{-1}$ is a $N \times N$ diagonal matrix with its non-zero entries as the inverse of weighted node strength $s_i$ (if $s_i=0$, then set ${\bf S}^{-1}(i,i)=0$), ${\bf A}={\bf S}^{-1}\hat{\bf A}=\{a_{ij}\}$ is the $N \times N$ network matrix with conformity behavior, and is in general non-symmetric due to weighted quantity ${\bf S}^{-1}$.

Input control signal ${\bf u}(\tau)$ is responsible for driving the node states of the network and the energy cost required is defined as \cite{lewis1995optimal,li2017control,li2017fundamental,duan2019target} 
$\mathcal{E}(T_f)=\sum\limits_{\tau=0}^{T_f-1} {\bf u}^T(\tau){\bf u}(\tau) $, where $T_f$ is the final control time, which is the allocated time that control signal ${\bf u}(\tau)$ must complete its tasks. Minimizing the energy cost, the discrete-time energy-optimal control signal can be derived as \cite{lewis1995optimal,duan2019target}: 
\begin{equation} \label{energy-optimal control signal}
{\bf u}^*(\tau)={\bf B}^T ({\bf A}^T)^{T_f-\tau -1}{\bf W}^{-1}({\bf x}_f-{\bf A}^{T_f}{\bf x}_0),
\end{equation}
where 
\begin{equation}\label{controllability gramian W}
{\bf W}(T_f)=\sum\limits_{\tau=0}^{T_f-1}{\bf A}^{T_f - \tau - 1}{\bf B}{\bf B}^T({\bf A}^T)^{T_f-\tau-1}
\end{equation}
is the $N \times N$ discrete-time controllability Gramian matrix, ${\bf x_0}=[x_1(0),x_2(0),...,x_N(0)]^T$ is the initial state vector and ${\bf x}_f=[x_1(T_f),x_2(T_f),...,x_N(T_f)]^T$ is the desired final state vector. Substituting ${\bf u}^*(\tau)$ into $\mathcal{E}(T_f)$, and assuming that ${\bf x}_0={\bf 0}$, the cost function becomes $\mathcal{E}(T_f)={\bf x}_f^T {\bf W}^{-1} {\bf x}_f$. In addition, to simplify the scope of the research, the final state vector ${\bf x}_f$ is normalized by restricting the Euclidean distance ${\bf x}^T_f {\bf x}_f$ to be one, leading to the normalized cost function $E(T_f)=\mathcal{E}(T_f)/{\bf x}^T_f {\bf x}_f$, which is bounded by the eigenvalues of the Gramian using the Rayleigh-Ritz theorem \cite{horn2012matrix}: 
\begin{equation}
\frac{1}{\abs{\eta^{\prime}_{\text{max}}({\bf W}(T_f))}}=\underline{E}^\prime \leq E(T_f) \leq \overline{E}^\prime=\frac{1}{\abs{\eta^{\prime}_{\text{min}}({\bf W}(T_f))}},
\end{equation}
where $\underline{E}^\prime (\overline{E}^\prime)$ is the lower (upper) bound of the energy cost, and $\abs{\eta^{\prime}_{\text{max}}({\bf W})}$ $(\abs{\eta^{\prime}_{\text{min}}({\bf W})})$ is the absolute maximum (minimum) eigenvalue of Gramian ${\bf W}$. In other words, to calculate the lower and upper bound energy costs, it is crucial to compute the inverse of the absolute maximum and minimum eigenvalues of the controllability Gramian.

The controllability Gramian exists in simplified form, which has an analytical expression comprising the complex network's eigenvalues. This can be obtained through eigendecompositions ${\bf A}={\bf P} {\bf D} {\bf P}^{-1}$ and ${\bf A}^T={\bf V}{\bf D} {\bf V}^{-1}$, where ${\bf D}_{ii}=\lambda_i$ is the $N \times N$ diagonal matrix containing eigenvalues of ${\bf A}$ $({\bf A}^T)$, sorted in ascending order of magnitude $|\text{Re}\lambda_1|\leq |\text{Re}\lambda_2| \leq ... \leq |\text{Re}\lambda_N|$, and ${\bf P}$ $({\bf V})$ is the $N \times N$ associated eigenvectors of ${\bf A}$ $({\bf A}^T)$. Substituting the eigendecompositions into Eqn.\ (\ref{controllability gramian W}), the Gramian, expressed in Hadamard product form \cite{yan2012controlling,yan2015spectrum,duan2019energy,duan2019target}, becomes:
\begin{equation}
{\bf W}(T_f)={\bf P} \sum\limits_{\tau =0}^{T_f-1}{\bf D}^{\tau}{\bf P}^{-1}{\bf BB}^T{\bf V D}^{\tau} {\bf V}^{-1}={\bf PMV}^{-1},
\end{equation}
where {\bf M}($T_f$) is the simplified controllability Gramian, with $(i,j)$ elements which can be obtained as (see supplementary information \cite{SIref})
\begin{equation} \label{M_ij}
{\bf M}_{ij}={\bf Q}_{ij} \frac{1-(\lambda_i \lambda_j)^{T_f}}{1-\lambda_i \lambda_j},
\end{equation}
with
\begin{equation} \label{Q_ij}
{\bf Q}_{ij}=[{\bf P}^{-1}{\bf BB}^T {\bf V}]_{ij}.
\end{equation}

Therefore, to study the $T_f$ scaling behaviors of normalized energy cost bounds, it suffices to analyze the eigenvalues of ${\bf M}(T_f)$ as $T_f$ varies:
\begin{equation} \label{energy cost bounds}
\frac{1}{\abs{\eta_{\text{max}}({\bf M}(T_f)) }}=\underline{E} \leq E(T_f) \leq \overline{E}= \frac{1}{\abs{\eta_{\text{min}}({\bf M}(T_f))}},
\end{equation}
where $\underline{E}$ $(\overline{E})$ is the lower (upper) bound of the energy cost, and $|\eta_{\text{max}}({\bf M}(T_f))|$ $(|\eta_{\text{min}}({\bf M}(T_f))|)$ is the absolute maximum (minimum) eigenvalue of ${\bf M}(T_f)$. 

While most eigenvalues are computed numerically, the eigenvalues of ${\bf M}(T_f)$ can be estimated analytically with the formulae \citep{duan2019energy,lam2011estimates}:
\begin{equation} \label{lower bound}
\abs{\eta_{\text{max}}({\bf M}(T_f))} \approx \abs{\text{Re} \Big[f(\overline{\alpha},\overline{\beta})\Big]}
\end{equation}
and
\begin{equation} \label{upper bound}
\abs{\eta_{\text{min}}({\bf M}(T_f))}\approx \abs{\text{Re} \frac{1}{f(\underline{\alpha},\underline{\beta})}},
\end{equation}
where $f(\alpha,\beta)=\sqrt{\frac{\alpha}{N}+\sqrt{\frac{N-1}{N}(\beta-\frac{\alpha^2}{N})}}$, and 
\begin{equation}\label{alpha overline}
\overline{\alpha}=\text{trace}({\bf M}^2),
\end{equation}
\begin{equation}\label{beta overline}
\overline{\beta}=\text{trace}({\bf M}^4),
\end{equation}
\begin{equation}\label{alpha underline 2}
\underline{\alpha}=\text{trace}[({\bf M}^{-1})^2],
\end{equation}
and 
\begin{equation}\label{beta underline}
\underline{\beta}=\text{trace}[({\bf M}^{-1})^4]
\end{equation}
are pertinent estimation parameters for obtaining analytical $\underline{E}$ or $\overline{E}$.

In subsequent analyses, through the use of Eqns.\ (\ref{lower bound}) and (\ref{upper bound}), the analytical scaling behaviors of the energy bounds with respect to $T_f$ are given in terms of small and large $T_f$ regime, and different number of driver nodes. Where possible, analytical estimations for $\underline{E}$ and $\overline{E}$ are made, if not, then approximate, or numerical ones, which are all validated against numerical computations of $\underline{E}$ and $\overline{E}$. 

\section{Results}\label{Results}
In this section, for the first time, the energy cost scaling laws of a complex network with discrete-time linear dynamics and conformity behavior is derived and quantified using the estimation formulae Eqns.\ (\ref{lower bound}) and (\ref{upper bound}). Necessarily, because the energy cost is somehow related to the network's eigenvalues, a discussion about the properties of the eigenvalues with and without conformity, and how they relate to system dynamics is given. Next, the energy costs needed to control complex networks with and without conformity are compared. Finally, the controllability properties of a SF network is studied, where it is showed, for the first time, that the scaling exponent $\gamma$ of a SF network tunes its controllability.

\begin{figure}[h!] 
\begin{center}
\includegraphics[width=1.0\linewidth]{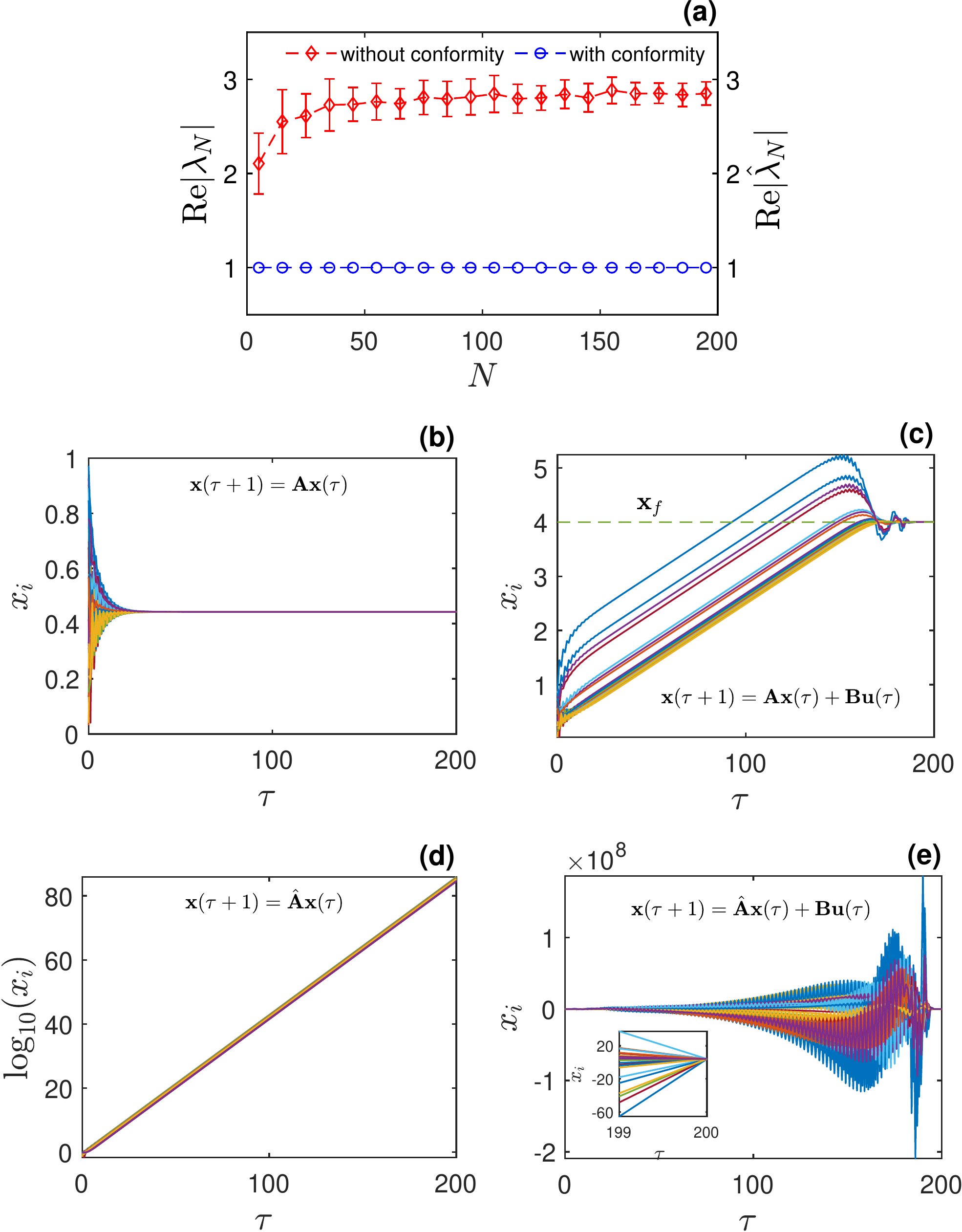}
\end{center}
\caption{Comparison of non-conformity and conformity discrete LTI models. The magnitudes of the largest eigenvalues of $\hat{\bf A}$ and ${\bf A}$ denoted by Re$|\hat{\lambda}_N|$ and Re$|\lambda_N|$ respectively for random network with fixed $\langle k \rangle=4$ and varying $N$ is shown in $(a)$, where each data point is computed as the mean (error bars are standard deviation) of $20$ independent network realizations. State equation evolving in $\tau$ with conformity dynamics, absent control ${\bf u}(\tau)$ is shown in (b), while (c) shows the node states being driven by one driver node towards the desired ${\bf x}_f={\bf 4}$. (d) depicts the state equation of non-conformity dynamics, which is unstable and diverges in $\tau$. (e) demonstrates the unstable system being driven by one driver node towards ${\bf x}_f={\bf 4}$, with inset showing the final two time step.} 
\label{fig:fig1}
\end{figure}

The energy cost bounds are estimated analytically using Eqns.\ (\ref{lower bound}) and (\ref{upper bound}), which are themselves dependent on Eqns.\ (\ref{alpha overline})\textemdash(\ref{beta underline}), whose calculations depend on the traces of ${\bf M}$ and ${\bf M}^{-1}$. Recalling Eqn.\ (\ref{M_ij}), ${\bf M}_{ij}$ is dependent on $\lambda_i$ (and $\lambda_j$), the eigenvalues of network ${\bf A}$, which is why it is important to quantify $\lambda_i$, as the energy cost bounds are coupled to them. As an example, using undirected Erdős–Rényi (ER) random networks with average degree $\langle k \rangle=4$ varying from system size $N=5$ up to $N=195$, it is shown in Fig.\ \ref{fig:fig1}(a), that when the system has conformity behavior, $|\text{Re}\lambda_i|<1$ for $i=1,2,...,N-1$, with $|\text{Re}\lambda_N|=1$. In contrast, for the same network structure, but without conformity behavior, let the system dynamics be:
\begin{equation}\label{LTI without conformity}
{\bf x}(\tau +1)= \hat{\bf A}{\bf x}(\tau) + {\bf B}{\bf u}(\tau),
\end{equation}
where $\hat{\bf A}$ is the network structure without conformity, with eigenvalues $\hat{\lambda}_i$ (for $i=1,2,...,N$), then it can be seen from Fig.\ \ref{fig:fig1}(a) that $\hat{\bf A}$ has maximum eigenvalue $|\text{Re}\hat{\lambda}_N|>1$, which makes the discrete-time system unstable because the node states diverge in time. In practice, an unstable system is undesirable and stability is the basic requirement for control \cite{chen1999linear}. Thus, when the discrete-time system has conformity, the network's eigenvalues change and become less than or equal to one, causing the system to be stable and controllable.

To demonstrate system stability and instability, the evolutions of node states in an ER network with $N=25$, and $\langle k \rangle=4$, with and without conformity are plotted in Fig.\ \ref{fig:fig1}(b) and (d), with $\bf{x}_0$ drawn from random uniform $[0,1]$. In Fig.\ \ref{fig:fig1}(b), when each node $i$ adapts its opinion as the average of its $n_i$ nearest neighbors, over time, consensus is reached within the network. On the other hand, when there is no conformity behavior, it can be seen in Fig.\ \ref{fig:fig1}(d) that opinions diverge in $\tau$, owing to Re$|\hat{\lambda}_N|>1$.

Regardless of system stability, both types of networks can be (numerically) controlled with one driver node. Setting ${\bf x}_f=[4,4,...,4]^T$, and using energy-optimal control signal $\bf{u}^*(\tau)$, with $\hat{{\bf A}}$ replacing $\bf{A}$ in (\ref{energy-optimal control signal}) for controlling network $\hat{\bf A}$, it is shown in Figs.\ \ref{fig:fig1}(c) and (e) that the node states could be driven toward ${\bf x}_f$. However, due to system instability, control signal ${\bf u}(\hat{\bf A},\tau)$ has to work much harder to bring the state vector towards the desired ${\bf x}_f$. At $T_f=8N$, the numerical energy cost difference between controlling the system with and without conformity is about $15$ orders of magnitude. While both conformity and non-conformity systems can be numerically controlled, the node states of the unstable system reach at least several orders of magnitude, which may in practice be infeasible.

Next, keeping in mind that $\lambda_i \leq 1$, the bounded estimates for the energy cost needed in controlling networks with discrete-time conformity-based linear dynamics is presented in the sections that follow. Before proceeding, it is worth mentioning that for discrete-time dynamical systems, $T_f$ cannot be below a certain value, $\underline{T_f}$, otherwise the controllability Gramian is not invertible, and the system is not controllable. Based on numerical experiments, it was found that when using $N$ drivers, $\underline{T_f}=1$, otherwise,
\begin{equation}
\underline{T_f} = \lceil \frac{N}{M} \rceil +1,
\end{equation} 
where $M$ is the number of control signals or driver nodes. 

\subsection{Lower bound $\underline{E}$} \label{results - lower bound}

\begin{figure*}
\begin{center}
\includegraphics[width=0.925\linewidth]{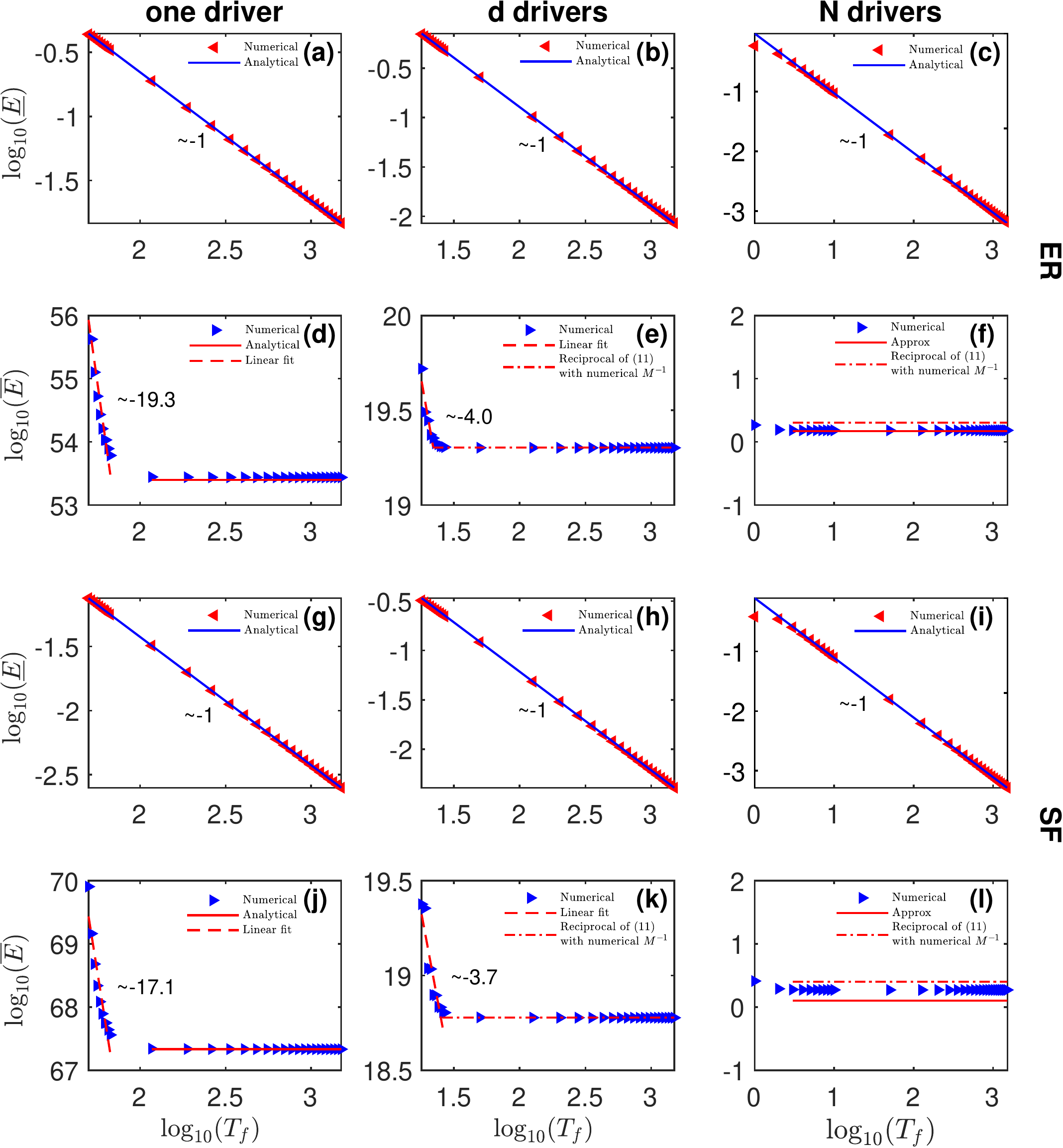}
\end{center}
\caption{Energy cost lower and upper bound for controlling model networks. Top-half (a)$\textendash$(f) and bottom-half (g)$\textendash$(l) panels relate to ER and SF networks (both are $N=50$, $\langle k \rangle=6$) respectively. Each column represents the same number of drivers, and $d=3$. Triangles are numerical calculations of the inverse of maximum and minimum absolute eigenvalues of ${\bf M}$, while solid lines, dashed lines, and dash-dotted lines are respectively the analytical scaling laws, numerical linear fit, and $\overline{E}$ estimates using (\ref{upper bound}), calculated from numerical ${\bf M}^{-1}$.} 
\label{fig:fig2}
\end{figure*}

\begin{figure*}
\begin{center}
\includegraphics[width=0.925\linewidth]{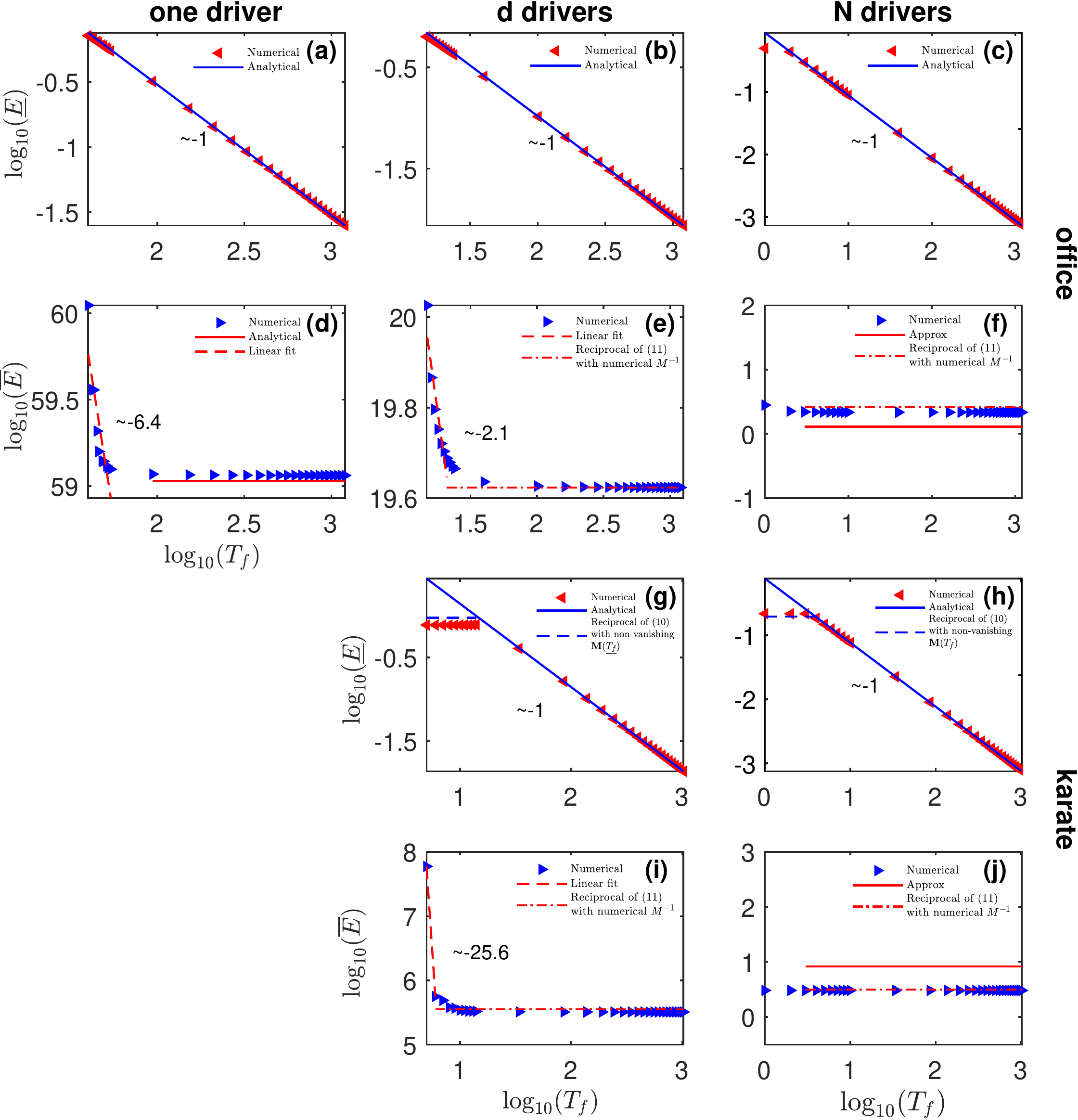}
\end{center}
\caption{Energy cost lower and upper bound for controlling real networks. Top-half (a)$\textendash$(f) and bottom-half (g)$\textendash$(j) panels relate to office and karate networks ($N=40$, $\langle k \rangle=11.9$, and $N=34$, $\langle k \rangle=4.6$) respectively. Each column represents the indicated number of drivers, with $d=3$ for office network and $d=10$ for karate network. There are no one driver node plots for karate network, because its minimum number of drivers \cite{yuan2013exact} is $N_D=10$. Triangles are numerical calculations of the inverse of maximum and minimum absolute eigenvalues of ${\bf M}$, while solid lines, red dashed lines, blue dashed lines, and dash-dotted lines are respectively the analytical scaling laws, numerical linear fit, small $T_f$ analytical $\underline{E}$, and $\overline{E}$ estimates using (\ref{upper bound}), calculated from numerical ${\bf M}^{-1}$.} 
\label{fig:fig3}
\end{figure*}

As indicated in Fig.\ \ref{fig:fig1}(a), networks ${\bf A}$ which have conformity behavior all have $|\text{Re}\lambda_N|=1$, with the rest of the eigenvalues $\lambda_i<1$ for $i=1,2,...,N-1$. Therefore, since ${\bf M}_{ij}={\bf Q}_{ij}\frac{1-(\lambda_i \lambda_j)^{T_f}}{1-\lambda_i \lambda_j}$, all $(\lambda_i \lambda_j)^{T_f}$ terms vanish in the large $T_f$ limit, with the exception of the last row/col element (using L'hopital's rule to evaluate the limit) ${\bf M}_{NN}=\lim \limits_{\text{Re}\lambda_N^2 \rightarrow 1}{\bf Q}_{NN} \frac{1-\text{Re}(\lambda_N^2)^{T_f}}{1-\text{Re}(\lambda_N^2)}={\bf Q}_{NN}T_f$. In the large $T_f$ limit, all ${\bf M}_{ij}$ values are small relative to ${\bf M}_{NN}$, which dominates, and applying (\ref{lower bound}), 
\begin{equation} \label{lowerbound generic}
\underline{E}\approx \text{Re} | \frac{1}{{\bf Q}_{NN}T_f} | \sim T_f^{-1},
\end{equation}
regardless of number of drivers, which is validated with model networks (ER and SF) as shown in Fig.\ \ref{fig:fig2}, and real networks (office \cite{bernard1979informant} and karate \cite{zachary1977information}) as shown in Fig.\ \ref{fig:fig3}. Note that the scaling $\underline{E}\sim T_f^{-1}$ is true for large $T_f$, but not necessarily true for small $T_f$. For small $T_f$, analytical $\underline{E}$ should be estimated with Eqns.\ (\ref{lower bound}), (\ref{alpha overline}) and (\ref{beta overline}) using the non-vanishing ${\bf M}_{ij}$. However, based on Figs.\ \ref{fig:fig2} and \ref{fig:fig3}, $\underline{E}\sim T_f^{-1}$ approximates the small $T_f$ lower bound energy cost with good agreement between analytical and numerical results in most cases. For Fig.\ \ref{fig:fig3}(g), the deviation of the numerical computations from the analytical scaling is noticeable, and the small $T_f$ analytical \underline{E} should be estimated with the non-vanishing ${\bf M}_{ij}$ instead.

The driver nodes placement is encoded in ${\bf Q}$ (Eqn.\ (\ref{Q_ij})), which affects the simplified controllability Gramian ${\bf M}$ (Eqn.\ (\ref{M_ij})), and is responsible for shifting the $\underline{E}$ plots vertically. When controlling the network with a single control signal attached to node $h$, ${\bf B}$ is a $N \times 1$ matrix, with entry $b_{h1}=1$, and the rest of the elements $b_{ij}=0$, leading to ${\bf B}{\bf B}^T={\bf J}^{hh}$, where ${\bf J}^{hh}$ is a single-entry matrix \cite{petersen2012matrix}, and substituting into Eqn.\ (\ref{Q_ij}), ${\bf Q}_{ij}=[{\bf P}^{-1}]_{jh}v_{hi}$. When using $d$ number of drivers (where $1<d<N$), ${\bf B}$ is a $N \times d$ matrix with the driver nodes placement $b_{ij}=1$ if the $j$-th control signal attaches to node $i$ and zero otherwise, leading to ${\bf B}{\bf B}^T={\bf J}^{d_1 d_1} + {\bf J}^{d_2 d_2} + ... + {\bf J}^{d_d d_d}$, where $d_k$ represents driver node $k$ for $k=1,2,...,d$, then ${\bf Q}_{ij}=\sum\limits_{k=1}^{d}[{\bf P}^{-1}]_{id_k}v_{d_k j}$. Finally, when controlling with all $N$ drivers such that all $N$ nodes each receive a control signal, ${\bf B}$ is a $N \times N$ matrix with $b_{ii}=1$ for $i=1,2,...,N$ and zero everywhere else, leading to ${\bf B}={\bf I}$, and ${\bf Q}_{ij}=[{\bf P}^{-1} {\bf V}]_{ij}$. Consequently, substituting the resultant ${\bf Q}_{ij}$ terms into Eqn.\ (\ref{lowerbound generic}),
\begin{equation} \label{lower bound piece wise}
\underline{E}\approx\begin{cases}
| \text{Re}[[{\bf P}^{-1}]_{Nh}v_{hN}T_f]^{-1} |, & \text{one driver,}  \\
| \text{Re}[\sum\limits_{k=1}^{d}[{\bf P}^{-1}]_{Nd_k}v_{d_k N}T_f]^{-1}|, & d \text{ drivers,} \\
| \text{Re}[[{\bf P}^{-1}{\bf V}]_{NN}T_f]^{-1}|,  & N \text{ drivers}.
\end{cases}
\end{equation}
Comparing Figs.\ \ref{fig:fig2}(a)\textendash(c), (g)\textendash(i), and Figs.\ \ref{fig:fig3}(a)\textendash(c), (g)\textendash(h), it can be seen that an increase in number of drivers leads to a decrease in $\underline{E}$. 

\subsection{Upper bound $\overline{E}$} \label{results - upper bound}
\begin{figure}[h!] 
\begin{center}
\includegraphics[width=1.0\linewidth]{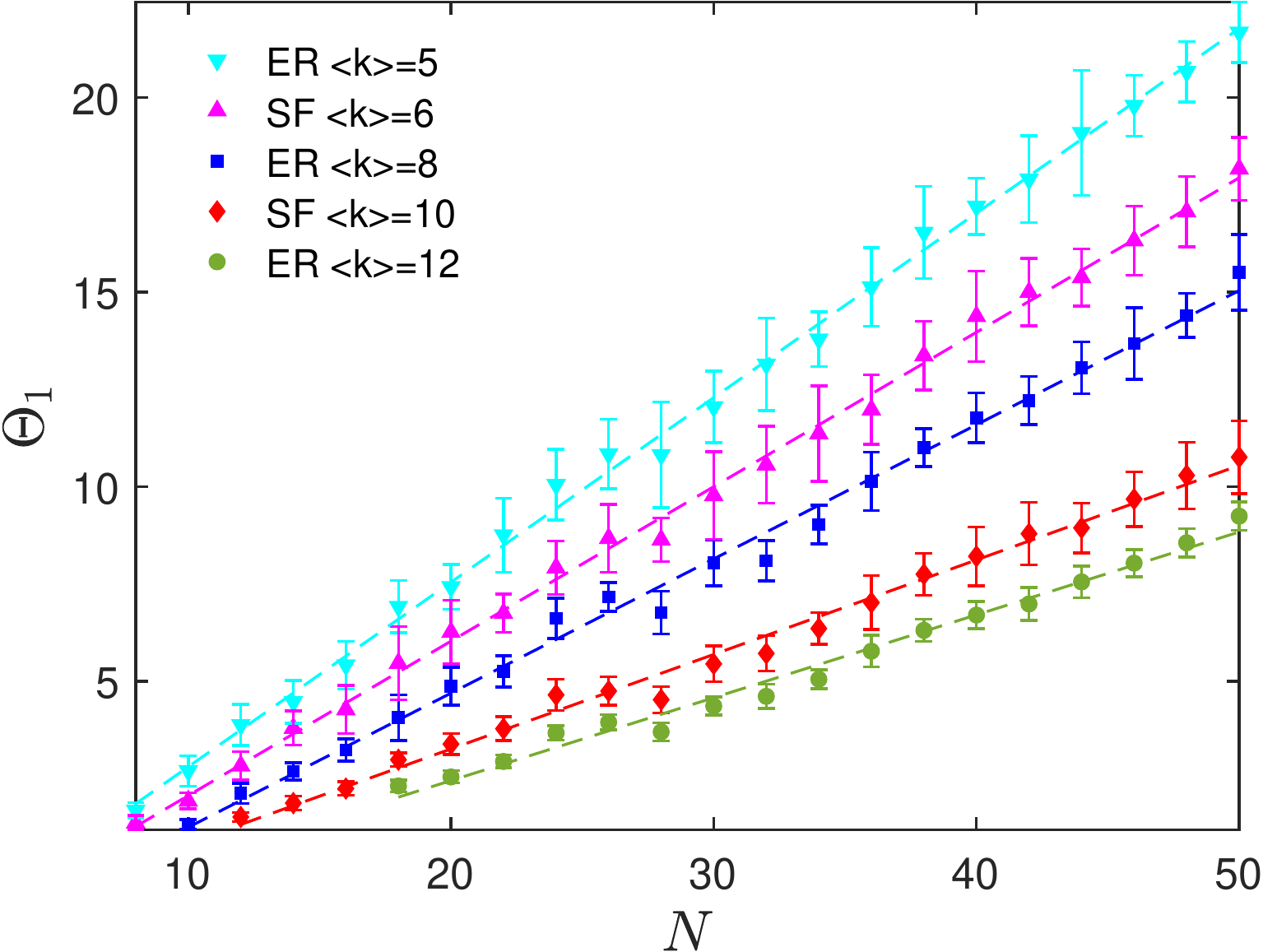}
\end{center}
\caption{$\Theta_1$ calculated from numerical linear fit in small $T_f$ regime of the one driver scaling law, $\overline{E}\sim T_f^{-\Theta_1}$, for varying $N$ and $\langle k \rangle$ with random and scale-free network topologies. Each data point is the mean of $20$ independent network realizations and error bar is the standard deviation. } 
\label{fig:fig4}
\end{figure}

Based on observing the numerical calculations of $\overline{E}$, there are two distinct regimes: small $T_f$ regime, characterized by linear scaling behavior $\overline{E}\sim T_f^{-\Theta}$, where $\Theta$ is a numerical value to be found, and large $T_f$ regime, where upper bound $\overline{E}$ converge to a constant value. Using different number of driver nodes leads to different scaling exponents such that $\overline{E}\sim T_f^{-\Theta_1}$, $\overline{E}\sim T_f^{-\Theta_d}$, and $\overline{E}\sim T_f^{-\Theta_N}$ when using one driver, $d$ drivers, and $N$ drivers respectively. Note that the range of small $T_f$ regime is relatively small and decreases with increasing number of driver nodes. Small $T_f$ regime is characterized by the linear scaling behavior $\overline{E}\sim T_f^{-\Theta}$, and not by the range of $T_f$. A simple algorithm can be used to perform the linear fitting of numerical $\overline{E}$: 
\begin{enumerate}
\item Calculate numerical $\overline{E}$ at $T_f=\{\underline{T_f},\underline{T_f}+1,\underline{T_f}+2,...,\underline{T_f}+9\}$, where $\underline{T_f}$ is the minimum permissible $T_f$
\item Linear fit the first two data points computed at $\underline{T_f}$ and $\underline{T_f}+1$ and accept this range as the valid small $T_f$ linear scaling regime
\item Linear fit the first three data points computed at $\underline{T_f}$, $\underline{T_f}+1$, and $\underline{T_f}+2$, and measure the $R$-squared value. If $R$-squared value $\geq 0.85$, accept the data points as the valid small $T_f$ linear scaling regime and move to step $4.$, else terminate
\item Repeat step $3$.\ with increasing number of data points: linear fit $\overline{E}$ computations evaluated at $T_f=[\underline{T_f},\underline{T_f}+i]$, for $i=3,4,...,9$.
\end{enumerate}
From this simple algorithm, the small $T_f$ regime is linearly fitted, where for one driver node calculations, small $T_f$ regime spans several data points, which decreases as number of driver nodes increases. Because using one driver node requires the most control energy \cite{yan2015spectrum}, it can be seen that $\Theta_1 > \Theta_d > \Theta_N $ when comparing Figs.\ \ref{fig:fig2}(d)\textendash(f), (j)\textendash(l), and Figs.\ \ref{fig:fig3}(d)\textendash(f), (i)\textendash(j). $\Theta_N$ is in general $<1$ and essentially negligible in $T_f$. In addition, $\Theta_1 \sim \frac{N}{ \langle k \rangle}$ regardless of network topology, as shown in Fig.\ \ref{fig:fig4}.

Next, $\overline{E}$ is derived for large $T_f$ limit, when controlling the network with one driver node, where a control signal is attached to node $h$, leading to input matrix ${\bf B}$ and simplified controllability Gramian matrix ${\bf M}$ the same as before in Section.\ \ref{results - lower bound}. For the upper bound, analytical $\overline{E}$ in the large $T_f$ regime has to be estimated using Eqns.\ (\ref{upper bound}), (\ref{alpha underline 2}), and (\ref{beta underline}), which are dependent on the inverse of the simplified controllability Gramian, ${\bf M}^{-1}$. While ${\bf M}$ has an analytical expression as shown in Eqn.\ (\ref{M_ij}), analytical ${\bf M}^{-1}$ has to be derived by applying ${\bf M}^{-1}=\frac{{\bf M}^*}{|{\bf M}|}$, where ${\bf M}^*$ is the adjoint matrix, and $|{\bf M}|$ is the determinant. Using Gaussian elimination to obtain the general expressions of ${\bf M}^*$ and $|{\bf M}|$ based on ${\bf M}$, analytical ${\bf M}^{-1}$ in the large $T_f$ limit is obtained as shown in Appendix (\ref{M inverse}). Note that (\ref{M inverse}) differs non-trivially from Ref.\ \cite{duan2019target}'s discrete-time system ${\bf M}^{-1}$, not just in terms of eigenvectors, but also in terms of form. Inspecting (\ref{M inverse}), ${\bf M}^{-1}$ elements lying in the last row/col are negligible when $T_f$ is large. Ignoring those terms and applying Eqn.\ (\ref{alpha underline 2}), analytical $\underline{\alpha}$ (\ref{alpha underline}) is a constant. Similarly, $\underline{\beta}$ can be obtained through Eqn.\ (\ref{beta underline}). Substituting $\underline{\alpha}$ and $\underline{\beta}$ into Eqn.\ (\ref{upper bound}), analytical $\overline{E}$ is a constant.

When using more than one driver ($d$ and $N$ drivers) to control the complex system, it is difficult to obtain analytical ${\bf M}^{-1}$ in the large $T_f$ limit. However, large $T_f$ analytical $\overline{E}$ can still be estimated with Eqns.\ (\ref{upper bound}), (\ref{alpha underline 2}), and (\ref{beta overline}) using ${\bf M}^{-1}$ computed numerically. From Figs.\ \ref{fig:fig2} and \ref{fig:fig3}, the $\overline{E}$ scaling behavior of the system is noted to be the same as when using one driver node, and $\overline{E}$ converge to a constant value in the large $T_f$ limit. In addition, for $N$ drivers, $\overline{E}$ can also be approximated with (\ref{upper bound}) using an approximate analytical ${\bf M}^{-1}$, which is a diagonal matrix such that its non-zero entries ${\bf M}^{-1}(i,i)=[{\bf M}_{ii}]^{-1}=\frac{1-\lambda_i^2}{[{\bf P}^{-1}{\bf V}]_{ii}}$. This approximation can be verified against numerical computation of ${\bf M}^{-1}$, where it is noted that the main diagonal elements are large in comparison to the rest of matrix elements. Applying approximate ${\bf M}^{-1}$ to Eqns.\ (\ref{alpha underline 2}) and (\ref{beta underline}), $\underline{\alpha}\approx\sum\limits_{i=1}^{N-1}\Big[\frac{1-\lambda_i^2}{[{\bf P}^{-1} {\bf V}]_{ii}}\Big]^2$ and $\underline{\beta}\approx\sum\limits_{i=1}^{N-1}\Big[\frac{1-\lambda_i^2}{[{\bf P}^{-1} {\bf V}]_{ii}}\Big]^4$, which when applied to (\ref{upper bound}), also yields a constant in the large $T_f$ limit. The validity of the analytical or approximate results are demonstrated in Figs.\ \ref{fig:fig2} and \ref{fig:fig3}. 

Regardless of network topology, it can be seen in Figs.\ \ref{fig:fig2} and \ref{fig:fig3} that the scaling behaviors of the upper bound $\overline{E}$ of the energy cost with respect to $T_f$ are very similar. 
\begin{equation}\label{upperbound piece wise}
\overline{E}\begin{cases}
\sim T_f^{-\Theta}, & \text{small }T_f, \\
\approx \text{constant}, & \text{large }T_f \text{ , one driver,} \\
\approx \text{constant}, & \text{large }T_f \text{ , $d$ drivers,}\\
\approx \text{constant}, & \text{large }T_f \text{ , $N$ drivers},\\
\end{cases}
\end{equation}
where in the small $T_f$ regime, because control signal ${\bf u}(\tau)$ has to drive the node states in a short amount of time, most energy is needed. Relaxing $T_f$ requirement on the driver nodes reduces $\overline{E}$ at a scaling rate of $\overline{E}\sim T_f^{\Theta_1}$, $\overline{E}\sim T_f^{\Theta_d}$ or $\overline{E}\sim T_f^{\Theta_N}$ depending on the number of drivers used. However, increasing $T_f$ beyond a certain point, $\overline{E}$ converge to a constant value. As expected, because increasing the number of drivers reduces control energy \cite{yan2015spectrum}, $\overline{E}$ is highest when using just one driver node.

\begin{figure*}[tbp]
\begin{center}
\includegraphics[width=0.95\linewidth]{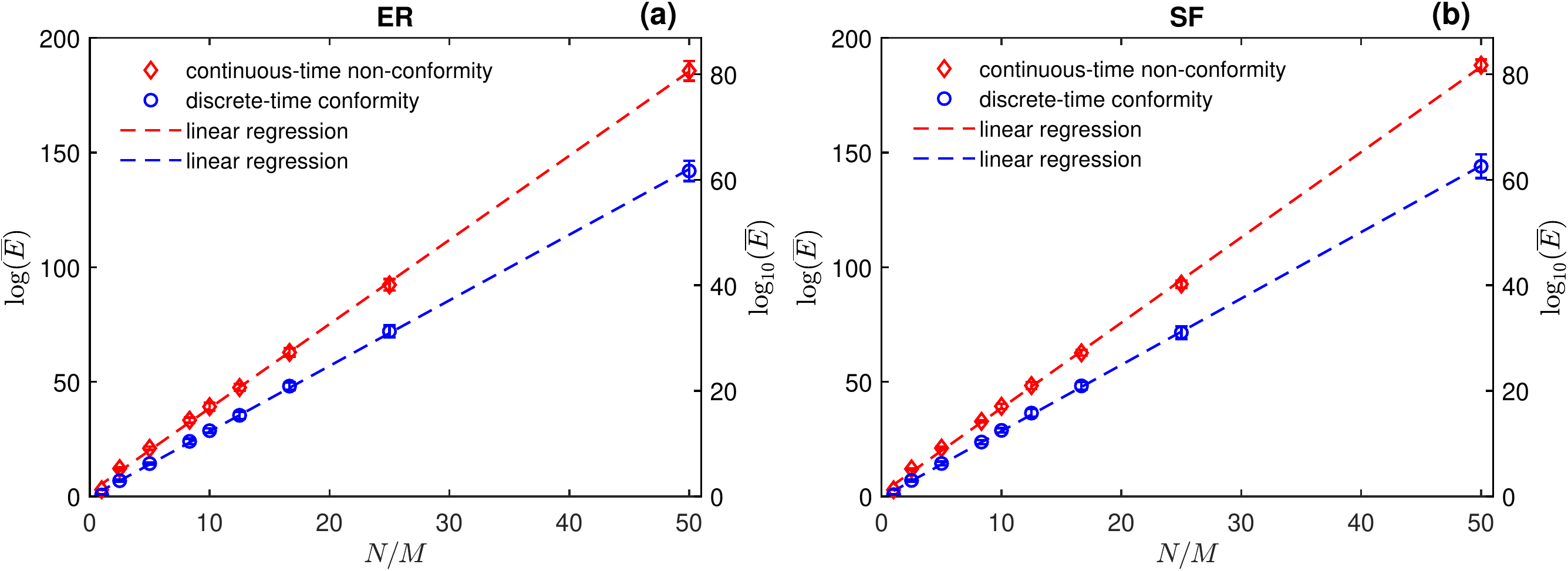}
\end{center}
\caption{Comparing the large $T_f$ upper bound energy cost $\overline{E}$ between the continuous-time non-conformity dynamical system and the discrete-time conformity-based dynamical system. The network sizes are all $N=50$, with average degree of $\langle k \rangle=8$. Both continuous-time non-conformity $\log(\overline{E})$ and discrete-time conformity $\log(\overline{E})$ can be linearly fitted against $N/M$, indicating that $\overline{E}\sim e^{N/M}$ for both types of systems. Overall, the conformity-based system requires less control energy. (a) shows $\overline{E}$ needed to control ER networks, while (b) shows $\overline{E}$ needed to control SF networks. Each data point is computed based on the mean of $20$ independent realizations, and error bars represent standard deviation. } 
\label{fig:fig5}
\end{figure*}

\subsection{Comparing large $T_f$ $\overline{E}$: conformity versus non-conformity}
From Figs.\ \ref{fig:fig2}, \ref{fig:fig3}, and Eqn.\ (\ref{upperbound piece wise}), it is apparent that $\overline{E}$ scales $\sim T_f^{-\Theta}$ in the small $T_f$ regime, eventually converging to a constant value in the large $T_f$ limit. This behavior is similar to those of continuous-time dynamics \cite{duan2019energy}, in particular, when the non-conformity continuous-time network $\tilde{\bf A}$ is not positive definite (PD, meaning that the eigenvalues of $\tilde{\bf A}$ are all positive) or positive semi-definite (PSD, meaning that the eigenvalues of $\tilde{\bf A}$ are all positive or zeros). A consequent natural question to ask is: How do the $\overline{E}$ values compare in the absence and presence of conformity? Answering this question could offer insights into the mechanism of conformity in the context of networked controllability. Before proceeding, it is worth clarifying why only large $T_f$ $\overline{E}$ values should be compared between the conformity-based discrete-time dynamical system and the non-conformity-based continuous-time dynamical system. Firstly, it is not sensible to compare conformity and non-conformity dynamics energy costs between the discrete-time systems (\ref{conformity behavior vectorised}) and (\ref{LTI without conformity}) because while the former is stable, the latter is unstable. Secondly, it is not sensible to compare the energy costs in the small $T_f$ regime, as the discrete-time dynamical system is limited by $\underline{T_f}$, while the continuous-time dynamical system can set $T_f$ to an arbitrarily small value. Thirdly, it is not sensible to compare lower bound, as conformity-based $\underline{E}$ goes to zero in the large $T_f$ limit. It should be noted that $T_f$ has different meaning in the discrete-time and continuous-time system; for example, $T_f=20$ is a relatively small $T_f$ for the discrete-time system, but $T_f=20$ is a large enough value for the continuous-time system such that further increase in $T_f$ would not lead to further reduction in energy cost \cite{yan2012controlling,duan2019energy}. For the remainder of this section, it can be assumed that the $T_f$ values for the continuous-time or the discrete-time dynamical system are chosen sensibly such that only the convergent large $T_f$ $\overline{E}$ values are being compared.

The continuous-time dynamical system has the following dynamics \cite{liu2011controllability}:
\begin{equation}\label{continuous-time LTI}
\dot{\bf x}(t) = \tilde{\bf A} \bf{x}(t) + {\bf B}{\bf u}(t),
\end{equation}
where $\tilde{\bf A}=\{\tilde{a}_{ij}\} \in \mathbb{R}^{N\times N}$ describes the network connection such that when node $j$ has a directed link to node $i$, $a_{ij}$ is non-zero, otherwise, $a_{ij}=0$, ${\bf B}=\{b_{ij}\} \in \mathbb{R}^{N \times M}$ accounts for where in the network the control signals are placed, for example, $b_{ij}=1$ if control signal $j$ attaches to node $i$, otherwise $b_{ij}=0$, $t$ is a continuous variable depicting instantaneous time, $\dot{\bf x}(t)=[\dot{x}_1(t),\dot{x}_2(t),...,\dot{x}_N(t)]^T$, ${\bf x}(t)=[x_1(t),x_2(t),...,x_N(t)]^T$, and ${\bf u}(t)=[u_1(t),u_2(t),...,u_N(t)]^T$ are respectively the instantaneous vector of rate of change of states, the state vector, and input control signals vector. Further, the continuous-time non-conformity network $\tilde{\bf A}$ is different from the discrete-time non-conformity network $\hat{\bf A}$ on the diagonal matrix elements. For the continuous-time system, to model system stability \cite{yan2015spectrum}, $\tilde{a}_{ii}= \sum\limits_{j=1}^N \tilde{a}_{ij} -\delta$, where $\delta$ represents a small perturbation to ensure system stability (in contrast, discrete-time $\hat{a}_{ii}=0$). For the simulations that follow, $\delta=0$, which is suitable for modelling opinion dynamics \cite{acemoglu2010spread}. 

The energy cost for the continuous-time dynamical system is defined to be \cite{rugh1996linear} $\mathcal{E}(T_f)=\int_{0}^{T_f}{\bf u}^T(t){\bf u}(t)dt$, whereupon optimizing \cite{yan2012controlling,duan2019energy}, leads to the continuous-time energy-optimal control signal
\begin{equation} 
{\bf u}^*(t)={\bf B}^T e^{\tilde{\bf A}^T (T_f - t)} {\bf W}^{-1}({\bf x}_f-e^{\tilde{\bf A}T_f}{\bf x}_0),
\end{equation}
where ${\bf W}=\int_{0}^{T_f} e^{\tilde{\bf A}(T_f -t)} {\bf B} {\bf B}^T e^{\tilde{\bf A}^T(T_f -t )} dt $ is the $N \times N$ continuous-time controllability Gramian, ${\bf x}_0 =[x_1(0),x_2(0),...,x_N(0)]^T$ is the initial state vector when $t=0$, and ${\bf x}_f=[x_1(T_f),x_2(T_f),...,x_N(T_f)]^T$ is the final state vector at $t=T_f$ that the complex system is being driven towards. Following analogous reasoning as Section.\ \ref{Problem formulation}, the bounds for the energy cost needed in controlling a continuous-time dynamical system is just the same as Eqn.\ (\ref{energy cost bounds}), except that $\eta_{\text{min}}$ and $\eta_{\text{max}}$ are the minimum and maximum eigenvalues of the simplified continuous-time controllability Gramian \cite{duan2019energy}:
\begin{equation} \label{M- continuous-time}
{\bf M}(T_f)=\tilde{\bf P}^T {\bf W} \tilde{\bf P},
\end{equation}
where $\tilde{\bf P}$ is the eigenvectors matrix of the eigen-decomposed continuous-time non-conformity matrix $\tilde{\bf A}=\tilde{\bf P} \tilde{\bf D} \tilde{\bf P}^T$, and ${\bf W}$ is the continuous-time controllability Gramian matrix. Thus, numerically, the upper bound energy cost is computed as:
\begin{equation} \label{upperbound continuous-time}
\overline{E}= \frac{1}{\abs{\eta_{\text{min}}({\bf M}(T_f))}}.
\end{equation}


Computing large $T_f$ $\overline{E}$ for the continuous-time non-conformity and the discrete-time conformity systems, their energy costs are compared in Fig.\ \ref{fig:fig5}. For both Figs.\ \ref{fig:fig5}(a) and (b), corresponding to ER networks and SF networks, $\log(\overline{E})$ is plotted on the vertical axes, while $N/M$, the total $N$ number of nodes of the complex network over the total $M$ number of control signals, is plotted on ther horizontal axes. For ease of reading the energy costs, the equivalent $\log_{10}(\overline{E})$ scale is also provided on the right-end vertical axes. Unsurprisingly, as noted from Fig.\ \ref{fig:fig5}, the linear relationship between $\log(\overline{E})$ and $N/M$ affirms the continuous-time results reported by Ref.\ \cite{yan2015spectrum}, that the upper bound $\overline{E}\sim e^{N/M}$, indicating that an increase in the number of drivers leads to a decrease in maximum control energy at an exponential rate. Different from what has already been reported in the literature, however, is that this scaling behavior, $\overline{E}\sim e^{N/M}$, is also true for the discrete-time dynamical system with conformity. Further, regardless of number of drivers and complex network topology, the upper bound energy cost $\overline{E}$ of the discrete-time conformity-based dynamical system always requires less control energy. When using only one control signal to drive the network such that $N/M=50$, the difference in $\overline{E}$ can be as much as $20$ orders of magnitude. The disparity in $\overline{E}$ lessens as the $M$ number of drivers increases. The results suggest that the mechanism of conformity always leads to a complex system that is easier to control, requiring less control energy.


\subsection{The role of scaling exponent $\gamma$ in the controllability of SF networks} \label{results - role of gamma}
When computing $N_D$ minimum number of drivers \cite{yuan2013exact} needed to ensure controllability of the karate network in Fig.\ \ref{fig:fig3}, it was found that $N_D =10$. Defining the degree of controllability as the fractional number of minimum drivers over the total number of nodes, $n_D=\frac{N_D}{N}$, and $\frac{\langle k \rangle}{N}$ as the density of connections of the network, the karate network was low in controllability (requiring high $N_d$ to be controlled), with $n_D\approx 0.29$, despite being dense in connections $\frac{\langle k \rangle}{N}\approx \frac{4.6}{34} \approx 0.14$. In contrast, as shown in Ref.\ \cite{wang2015controlling}, SF networks with $\gamma=3$ have $n_D$ decreasing monotonically with increasing $\langle k \rangle$ towards $n_D=\frac{1}{N}$, where starting at $\frac{\langle k \rangle}{N}\approx\frac{3}{500}=0.006$, are sufficiently dense and controllable with one driver. Therefore, with the karate network as a counterexample, $\langle k \rangle$ cannot be the only mechanism in determining the controllability of SF networks. 

\begin{figure}[h!] 
\begin{center}
\includegraphics[width=1\linewidth]{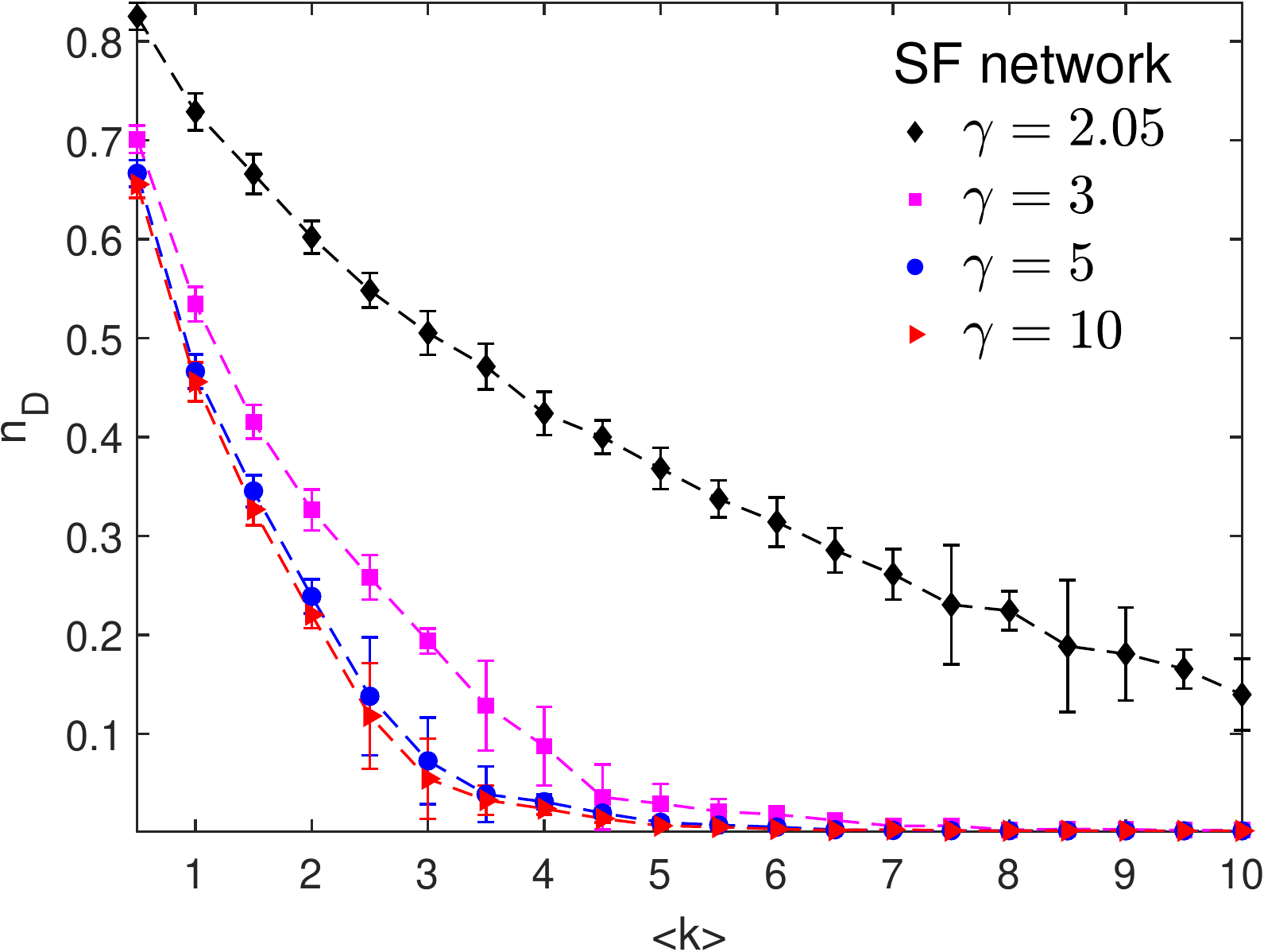}
\end{center}
\caption{The degree of controllability $n_D$ of SF networks with varying $\gamma$ and $\langle k \rangle$ values. Each data point is the mean of $20$ independent network realizations, with the error bar being the standard deviation.} 
\label{fig:fig6}
\end{figure}

The scaling exponent $\gamma$ also affects the controllability of SF networks. Using the static model \cite{goh2001universal} to generate SF networks with varying $\gamma$ (where $2<\gamma<\infty$), it is shown in Fig.\ \ref{fig:fig6}, for fixed $N=500$, that SF networks with lower $\gamma$ ($2.05$) are less controllable. On the other hand, high $\gamma$ ($5$ and $10$) does not improve the controllability of SF networks significantly when compared to $\gamma=3$. All scaling exponent $\gamma$ curves display decreasing $n_D$ with increasing $\langle k \rangle$, and the curves are the same without or without conformity dynamics.

\section{Discussion} \label{Discussion}

\begin{table}[b]
\centering
\captionof{table}{Summary of the scaling behaviors of the lower bound $\underline{E}$ and upper bound $\overline{E}$ of the energy cost with respect to $T_f$ in small and large $T_f$ regime using one, $d$, and $N$ number of drivers. } \label{tab:table summary} 
\begin{tabular}{ C{0.2in} C{0.6in} C{1in} C{.6in} C{.6in} }
\hline
\hline
& {$T_f$ regime}  & one driver & $d$ drivers & $N$ drivers  \\%
\hline
$\underline{E}$ & small/large & $\sim T_f^{-1}$ & $\sim T_f^{-1}$ & $\sim T_f^{-1}$  \\  
$\overline{E}$ & small & $\sim T_f^{-\Theta_1}$ & $\sim T_f^{-\Theta_d}$ & $\sim T_f^{-\Theta_N}$ \\
$\overline{E}$ & large & (\ref{upper bound}) with (\ref{M inverse}) & constant & constant  \\ 
\hline
\hline
\end {tabular}\par
\bigskip
\end{table}

In this paper, the energy cost needed to control a complex network with conformity behavior is studied. To do so, a series of scaling laws were derived, based on the number of driver nodes, to characterize the control energy as a function of $T_f$ in terms of its lower bound and upper bound, allowing for the energy cost to be estimated. A summary of the scaling laws can be found at Table \ref{tab:table summary}. In particular, it was found that lower bound scales as $\underline{E} \sim T_f^{-1}$, suggesting that in the large $T_f$ limit, no energy cost is needed by the driver nodes to steer the state vector towards the least costly direction. If the intended final state vector coincides with the network consensus opinion, which can be reached naturally with conformity dynamics, then no further action is required by the driver nodes. Upper bound energy cost decreases with increasing $T_f$ at a rate of $\overline{E}\sim T_f^{-\Theta}$ in the small $T_f$ limit, suggesting that in this regime, setting a relatively larger $T_f$ always leads to less energy cost. Further, it was found that $\Theta_1 \sim \frac{N}{\langle k \rangle}$ regardless of network topology, and the small $T_f$ regime scaling behavior depends only on system size and average node degree. In the large $T_f$ limit, $\overline{E}$ converges to a constant value, indicating that in this regime, the energy cost cannot be lowered any further by increasing $T_f$. Overall, these scaling laws have implications about the trade-off between final control time $T_f$ and energy cost and is useful to the design of control strategies in complex social networks with conformity behavior.

It is interesting to note the difference in controllability and energy cost of the networked systems with and without conformity. From the point of view of dynamical systems, the discrete-time non-conformity model $\hat{\bf A}$ is always practically uncontrollable owing to system instability. Yet, the same network structure with conformity in discrete-time linear dynamics, ${\bf A}={\bf S}^{-1}\hat{\bf A}$, is always stable and controllable, suggesting that the mechanism of conformity can make an otherwise unstable and uncontrollable discrete-time linear dynamical complex system stable and controllable. In terms of controllability, it was found that the fractional minimum number of drivers that guarantees the controllability of a SF network is unaffected by the mechanism of conformity. Instead, given the scale-free degree distribution $P(k)\sim k^{- \gamma}$ of a SF network, it was found that the scaling exponent $\gamma$ tunes the controllability, and heterogeneous (smaller $\gamma$) \cite{nacher2012dominating} SF networks are less controllable, requiring a higher minimum number of driver nodes. In terms of energy cost, networks with discrete-time linear dynamics and conformity behavior and networks with continuous-time linear dynamics but no conformity behavior both scale as a function of number of drivers as $\overline{E}\sim e^{N/M}$, indicating that regardless of conformity, adding more driver nodes always reduces the maximum control energy at an exponential rate. Nevertheless, comparing the convergent large $T_f$ $\overline{E}$ values of the different systems shows that the conformity-based model always requires less control energy with a difference of $\overline{E}$ values by as much as $20$ orders of magnitude when using one driver node to control networks of system size $N=50$. All in all, the results suggest that the mechanism of conformity plays an important role in making a complex social system more controllable.

Objectively speaking, the degree to which a social system is controllable has neutral connotations: It is neither better nor worse to be more/less controllable. For example, a complex social system that is more difficult to control is at the same time more resilient against malicious social manipulations as it is against positive social change. The converse is true for a complex social system that is easier to control. Thus, from the point of view of governance, conformity behavior in complex social systems should be discouraged in some situations, for example, the sharing of controversial or polarizing news, and encouraged in some others, for example, the propagation of ideas which can lead to positive social change such as recycling. Recently, a few socio-physics works have been successful in corroborating their mathematical models with real world data, such as the distributions of agree, disagree, or undecided in controversial and non-controversial topics using poll data \cite{krause2019repulsion}, and the emergence of echo chambers in social networks surrounding politically controversial topics using Twitter data \cite{baumann2020modeling}. It would be interesting to corroborate or even disprove, with real world data, the claim of the present research that conformity in social networks always leads to a situation where the network is easier to control or influence. Further, a limitation of networked controllability is that only linear dynamics can be studied \cite{liu2016control}, while various socio-physics models can be nonlinear or have states which are not continuous variables \cite{castellano2009statistical}. In these types of models, how would conformity dynamics influence the system as a whole?

\begin{acknowledgments}
The authors are grateful to Dr.\ Chew Lock Yue and Dr.\ Chiam Keng-Hwee for helpful discussion and suggestions, and Matthew Ho for guidance on creating high resolution figures from Matlab. H.C. and E.H.Y. acknowledge support from Nanyang Technological
University, Singapore, under its Start Up Grant Scheme (04INS000175C230).\end{acknowledgments}

\begin{appendix} 

\section{Methods}\label{appendix - methods}

System stability, network matrix diagonal entries and numerical precision require careful attention during computation. Unlike its continuous-time counterparts, which could be stabilized by introducing negative terms along the network diagonals \cite{yan2015spectrum}, the discrete-time system cannot be stabilized the same way. Introducing diagonal entries in $\hat{{\bf A}}$ would cause system instability for both non-conformity and conformity discrete-time dynamics. Therefore, in this work, for discrete-time dynamics, the diagonal entries of $\hat{\bf A}$ are zero, and conformity-based model is stable with eigenvalues $\lambda_i \leq 1$. When computing the state equation of the unstable non-conformity discrete-time dynamics, which diverges, high numerical precision is needed, which can be achieved using Advanpix \cite{mct2020}. Computing the unstable state equation without high numerical precision, using the standard double precision, would lead to erroneous calculations where they would wrongly stabilize in large $\tau$. Further, one driver node control typically requires high energy cost, and $\overline{E}$ has to be calculated with high numerical precision.

The model networks used in this paper are constructed with the standard Erdős–Rényi random network algorithm, and the static model \cite{goh2001universal}, where in Fig.\ \ref{fig:fig2}, $\gamma=2.5$, in Fig.\ \ref{fig:fig5}, $\gamma=3.0$, and in Fig.\ \ref{fig:fig6}, $\gamma$ is varied. All model networks $\hat{\bf A}$ and $\tilde{\bf A}$ are undirected, symmetric, and weighted, with link weights $\hat{a}_{ij}=\hat{a}_{ji}$ or $\tilde{a}_{ij}=\tilde{a}_{ji}$ drawn from random uniform $(0,1]$. The network construction algorithms do not guarantee that the graph object does not have isolated nodes, and for one driver node calculations, checks were done to ensure that each node has at least one connection, otherwise the network was discarded and replaced. For the real networks, the karate network is unweighted with link $\hat{a}_{ij}=1$ if nodes $i$ and $j$ are connected and zero otherwise, the office network is weighted but with integer $\hat{a}_{ij}$ greater or equal to $1$ if nodes $i$ and $j$ are connected and zero otherwise.

With sufficiently high average degree $\langle k \rangle$, the conformity-based model networks used have $N_D=1$, and any random choice of driver nodes (for $d$ drivers and one driver) would suffice to ensure network controllability. It should be noted that $k$, node degree, and $s$, node strength, are distinct from each other; where the former is calculated from the adjacency matrix, and the latter from $\hat{\bf A}$ whose links $\hat{a}_{ij}$ are weighted. For computing minimum driver nodes \cite{yuan2013exact}, the Matlab software written by Patel was used \cite{patel2015automated}.

\section{One driver node analytical equations} \label{appendix - analytical eqns}

\begin{widetext}
\begin{equation} \label{M inverse}
{\bf M}^{-1}(i,j)=\begin{cases}
\dfrac{(-1)^N \prod\limits_{k=1}^{N-1}(1-\lambda_i \lambda_k)\prod\limits_{\substack{l=1 \\ l\neq i}}^{N-1}(1-\lambda_l \lambda_i)}{[{\bf P}^{-1}]_{jh}v_{hi}\prod\limits_{\substack{k=1\\k\neq i}}^{N-1}(\lambda_i - \lambda_k) \prod\limits_{\substack{l=1\\l\neq i}}^{N-1}(\lambda_l - \lambda_i)},  & i=j\text{, }  (N-1) \times (N-1) \text{ block,}\\
\Vast[\dfrac{(-1)^{N+1}}{(\lambda_j-\lambda_i)v_{hi}[{\bf P}^{-1}]_{jh}}\Vast]\
\Vast[ \dfrac{\prod\limits_{k=1}^{N-1}(1-\lambda_i \lambda_k) \prod\limits_{\substack{k=1 \\ k \neq i}}^{N-1}(1-\lambda_k \lambda_j)}{\prod\limits_{\substack{k=1 \\ k \neq i}}^{N-1}(\lambda_i-\lambda_k)\prod\limits_{\substack{k=1 \\ k \neq j,i}}^{N-1}(\lambda_k-\lambda_j)} \Vast], & i\neq j\text{, } (N-1) \times (N-1) \text{ block,}\\
\dfrac{1}{q_{NN}T_f}=\dfrac{1}{[{\bf P}^{-1}]_{Nh}v_{hN}T_f},  & i=j=N, \\
\Vast[ \dfrac{(-1)^N}{[{\bf P}^{-1}]_{jh}v_{hi}(\lambda_j - \lambda_i)T_f}\Vast]\Vast[ \dfrac{\prod\limits_{k=1}^{N-1}(1-\lambda_i \lambda_k)\prod\limits_{\substack{l=1 \\ l \neq i}}^{N-1}(1-\lambda_l\lambda_j)}{\prod\limits_{\substack{k=1\\k\neq i}}^{N-1}(\lambda_i - \lambda_k)\prod\limits_{\substack{l=1 \\ l \neq j}}^{N-1}(\lambda_l - \lambda_j)} \Vast],& i \neq j \text{, } N^{\text{-th}} \text{ row/col.}
\end{cases}
\end{equation}

\begin{align} \nonumber \label{alpha underline}
\underline{\alpha}=&\text{trace}[({\bf M}^{-1})^2]\approx \sum\limits_{i=1}^{N-1} \sum\limits_{\substack{a=1 \\ a\neq i}}^{N-1}[{\bf M}^{-1}]_{ia}[{\bf M}^{-1}]_{ai}+[{\bf M}^{-1}]_{ii}[{\bf M}^{-1}]_{ii}\\
\approx &\sum\limits_{i=1}^{N-1} \sum\limits_{\substack{a=1 \\ a\neq i}}^{N-1} \Vast[\dfrac{1}{(\lambda_a-\lambda_i)v_{hi} [{\bf P}^{-1}]_{ah}}\Vast] \Vast[ \dfrac{\prod\limits_{k=1}^{N-1}(1-\lambda_i\lambda_k)\prod\limits_{\substack{k=1 \\ k\neq i}}^{N-1}(1-\lambda_k \lambda_a)}{\prod\limits_{\substack{k=1 \\ k \neq i}}^{N-1}(\lambda_i - \lambda_k) \prod\limits_{\substack{k=1 \\ k \neq a,i}}^{N-1}(\lambda_k-\lambda_a)}\Vast]\Vast[\dfrac{1}{(\lambda_i - \lambda_a)v_{ha}p^{-1}_{ih}} \Vast]\Vast[ \dfrac{\prod\limits_{k=1}^{N-1}(1-\lambda_a\lambda_k)\prod\limits_{\substack{k=1\\k\neq a}}^{N-1}(1-\lambda_k \lambda_i)}{\prod\limits_{\substack{k=1 \\ k\neq a}}^{N-1}(\lambda_a - \lambda_k)\prod\limits_{\substack{k=1 \\ k \neq i,a}}^{N-1}(\lambda_k - \lambda_i)}\Vast] \nonumber \\
&+ \Vast[ \dfrac{\prod\limits_{k=1}^{N-1}(1-\lambda_i\lambda_k)\prod\limits_{\substack{l=1\\l \neq i}}^{N-1}(1-\lambda_l \lambda_i)}{[{\bf P}^{-1}]_{ih}v_{hi} \prod\limits_{\substack{k=1 \\ k\neq i}}^{N-1}(\lambda_i - \lambda_k) \prod\limits_{\substack{l=1 \\ l\neq i}}^{N-1}(\lambda_l - \lambda_i)  } \Vast]^2 ,
\end{align}

See supplementary information for additional information \cite{SIref}.
\end{widetext}

\end{appendix}


\bibliography{ref}

\end{document}